\documentstyle[12pt,epsfig]{article}                                               
\newcommand{\simlt}  {\raisebox{-.6ex}{$\stackrel{\textstyle <}{\sim}$}}
\newcommand{\simgt}  {\raisebox{-.6ex}{$\stackrel{\textstyle >}{\sim}$}}
\begin{document}                                                                
\begin{flushright}
RAL-TR/1999-087 \\
24 December 2001 \\   
\end{flushright}                                                               
\vspace{0 mm}                                                                   
\begin{center}
{\Large
Nucleon Structure, Duality and
Elliptic Theta Functions}            
\end{center}
\vspace{5mm}                      
\begin{center}                      
{W. G. Scott\\                      
Rutherford Appleton Laboratory\\    
Chilton, Didcot, Oxon OX11 0QX. UK \footnotemark[1]} 
\end{center}
\vspace{1mm}
\begin{abstract}
\baselineskip 0.6cm
\noindent 
Nucleon structure functions 
are shown to have a qualitative 
(or `formal') relation to 
the classical elliptic theta functions.
In particular,
$\theta_1'/2$ shows a clear resemblance
to a non-singlet structure function like $xF_3$.
In the appropriate range,
the $Q^2$-dependence of the moments 
of $\theta_1'/2$ is in near-quantitative 
agreement with QCD,
and at low-$Q^2$ the moments
converge to a common value,
as observed empirically for $xF_3$.
At very high $Q^2$ ($Q \rightarrow M_{PL}$),
$\theta_1'/2 \rightarrow \eta^3(x)$
(where $\eta(x)$ is the Dedekind eta function)
while $xF_3$ in the same limit appears 
closer to $\eta^5(x)$.
A comparison
of the theta function identity 
$\theta_2^4(0)=\theta_3^4(0)-\theta_4^4(0)$
with the relation
$xF_3 = Q(x)-\bar{Q}(x)$
suggests that singlet structure functions
have more in common with $\theta_3$ and $\theta_4$.
The possibility of interesting 
large-$x$/small-$x$
`duality' relations for structure functions
emerges naturally from the analysis.
\end{abstract}
\footnotetext[1]{E-mail:w.g.scott@rl.ac.uk}
\newpage 
\baselineskip 0.6cm
 
Structure functions 
have long been recognised as fundamental
by experimentalists, 
phenomenologists and theorists alike,
and impressive progress has been 
made over the years in achieving a
precise, quantitative 
understanding of their behaviour,
over an increasingly large a kinematic range,
mainly within the context of perturbative QCD
(see for example Ref.\ \cite{HER}).
A quantitative description/representation
of structure functions
to match that achieved to date in perturbative QCD,
however,
is not the immediate (nor indeed the ultimate)
aim of the present paper,
and neither in fact is QCD
our point of departure here.
Rather the aim of this paper
is to draw attention
to a rather startling 
(if qualitative)
{\em empirically}-observed
similarity/relation
existing
between nucleon structure functions
and the classical elliptic 
(or Jacobi \cite{ELL}) theta functions,
almost certainly involving,
it would appear,
the very highest energy scales,
eg.\ the unification scale 
or the Planck-energy. 

Theta functions of various kinds,
and in particular the modular symmetry
inherent to theta functions,
are themselves recognised as fundamental,
mainly within the context 
of more-abstract theory,
most notably in string-theory \cite{GSW}.
Triggering the present paper,
and drawing strongly on considerations
of modular symmetry
(with specific reference to the Jacobi theta functions 
in the scale-invariant case)
are recent results \cite{WIT}
on strong/weak-coupling duality,
monopole condensation etc,
which are generally believed
to bear on the confinement problem,
and which in the long-term
might reasonably be expected
to lead to practical calculations 
of relevant non-perturbative quantities
in theories akin to QCD.
Thus while (as we shall see)
there do seem to be points of contact
from our observations to perturbative QCD
and to a number of associated 
phenomenological ideas/models,
we do have to draw attention
to the possibility of 
an important link
to what one might call 
`fundamental theory' here.
Naturally, we continue to emphasise
the empirical aspects
of our observations,
which we regard as striking 
and interesting in their own right,
regardless of explanation.

With duality ideas in mind therefore,
we begin by observing that the modular transformation
formula (Weyl inversion)
for a simple theta function $\theta[\tau]$ viz:
\begin{equation}
    \theta[\tau] = \sqrt{\frac{i}{\tau}}\theta[-1/\tau]
\end{equation}
is arguably just the kind of relation
one might hope to establish
between the strong-coupling ($\tau \rightarrow 0$)
and weak-coupling ($\tau \rightarrow \infty$)
domains in a theory like QCD. 
The parameter $\tau$ is in general
complex (with Im $\tau > 0$)
but for definiteness we focus here on 
the case that $\tau$ is purely imaginary:
\begin{equation}
   \hspace{2.5cm}  \tau = it \hspace{2cm} (t > 0)
\end{equation}
so that $t =-i\tau$ is real and positive
in this case
(via the notion of the `running' coupling,
we might think of $t$ as
the logarithm of some `energy', see below).

The four classical theta functions
$\theta_i(z|\tau)$, $i=1 - 4$ \cite{ELL} 
all have well-defined modular transformation
properties, each satisfying a relation 
similar to Eq.~1
in the case that $z=0$.
In Ref.\ \cite{WIT}
the functions
$\theta_2$, $\theta_3$ and $\theta_4$
(evaluated at $z=0$)
arise in connection
with the analysis of 
spin structures
in massless S-dual
theories with dimensionless couplings.
In this paper we
make particular use of the 
function $\theta_1'(z|\tau)$:
\begin{equation}
\theta_1'(z|\tau)=2 \sum_{n = 0}^{\infty} (-1)^n 
                e^{i2\pi \tau (2n+1)^2/8} (2n+1)\cos (2n+1)z  
\end{equation}
obtained from $\theta_1$ by differentiation
with respect to $z$
(the $\theta_i$ are by no means all independent, eg.\
$\theta_2(z)=\theta_1(z+\pi/2)$, $\theta_3(z)=\theta_4(z+\pi/2)$ 
and 
$\theta_1'(0)=\theta_2(0) \, \theta_3(0) \, \theta_4(0)$ \cite{ELL},
where $\theta_i(0)$ denotes a theta function
evaluated at $z=0$).

With regard to the experimental data 
on structure functions,
theoretical fits based on
perturbative QCD
(which makes rather clear-cut predictions
for the $Q^2$ dependence of structure functions,
at sufficiently high $Q^2$)
have typically been used to
interpolate and extrapolate
existing experimental data
to cover the full kinematic range
of current/future experimental interest
(ie.\ up to and including LHC energies).
By way of example, 
in Figure~1a we show (solid curves)
the $x$-dependence of the isospin-averaged,
non-singlet neutrino-nucleon structure function $xF_3$,
based on the MRS \cite{MRS} fit, 
for factor-of-ten increments in $Q^2$,
for $Q^2 = 1 - 10^7$ GeV$^2$
(here $Q^2$ is the square of the four-momentum transfer
and $x$ is the usual Bjorken scaling variable).
The data used in these fits
cover the range $Q^2 = 1 - 10^4$ GeV$^2$,
the fit being deliberately restricted
to $Q^2 > 1$ GeV$^2$, 
to reduce the risk of trespass
into the non-perturbative regime.
While the $x$-dependence at fixed $Q^2$
is largely determined by the data in these fits
(perturbative QCD has much less to say
about the expected $x$-dependence),
a number of phenomenological ideas/models
do serve as some guide in selecting
appropriate functional forms for fitting.
Regge-theory \cite{REG}, for example, suggests
an $x^{\frac{1}{2}}$ dependence for $xF_3$ at small $x$
and dimensional counting rules \cite{FAR} 
suggest $(1-x)^3$ for the behaviour as $x \rightarrow 1$,
these predictions being expected to hold 
at some not-too-large $Q^2$ 
($Q^2 \simeq 1$ GeV$^2$).
The sequence of curves
shown in Figure~1a prompts
the question of what is the high energy limit
of the structure function or 
(perhaps more relevantly)
what is the extrapolated form
at the unification scale or the Planck energy?

The undenied continued sucess of 
thermodynamic/statistical-models 
\cite{THERM} \cite{STAT}
of structure functions
would certainly suggest
that statistical-mechanical 
(or combinatorial) 
considerations could underlie
the form of the structure functions
at high energies 
(it has also been noted directly from experiment
that the $x$-dependence of structure functions becomes more 
exponential in character with increasing energy \cite{UA1}).
Closely associated with theta functions,
the Dedekind eta function $\eta[\tau]$
obeys the same modular transformation formula
as the simple theta function 
(Eq.~1 above)
and has a clear combinatorial significance (see below).
Furthermore the $x$-dependence of the eta function, given by:
\begin{equation}
\eta(x)= x^{\frac{1}{24}} \prod_{n=1}^{\infty} (1-x^n)
\end{equation}
($\eta(x) \equiv \eta[\tau]$, 
$x = \exp(2\pi i \tau)$)
is remarkably reminiscent
of the $x$-dependence expected
of a non-singlet structure function,
cf.\ the Regge prediction etc.\ (above).
With $t =-i\tau$ taken to be real and positive
as before (cf.\ Eq.~2) we have:
\begin{equation}
t=\frac{1}{2\pi} \ln \frac{1}{x}
\end{equation}
with $x$ also real and positive and 
satisfying $0 < x < 1$ as required.
With this
(perhaps unexpected) 
association of $t$  
with the scaled parton energy $x$
(rather than $Q=\sqrt{Q^2}$),
the modular transformation
(cf.\ Eq.~1) would become a 
large-$x$/small-$x$ relation,
ie.\ a symmetry versus
$\ln t = \ln \frac{1}{2\pi} \ln 1/x$
centered about a `critical' x-value 
$x_0 = \exp(-2\pi)$.

In Eq.~4
the pre-factor $x^{1/24}$
is known to be essential 
to the modular covariance.
The infinite product has a 
clear combinatorial significance
as the reciprocal of the generating function 
for the classical partition function \cite{RAM}.
Physically it is just the reciprocal 
of the ordinary partition function
for a system comprising a (freely) variable 
number of indistinguishable massless bosons 
moving freely in one dimension
(the index 1/24 in Eq.~4
is then interpreted 
as the associated Casimir 
(or vacuum) energy \cite{CASM}).
The generalisation to more than one
`colour' of boson
is readily obtained by raising the
generating function to the power of the
number of boson colours \cite{KAC}.
Functions like 
$\eta^3(x)$, $\eta^8(x)$ \dots 
appear naturally as 
operator traces or `characters' in relation to
the infinite dimensional lie algebras
$\hat{su}_2$, $\hat{su}_3$ \dots 
which are
equivalent to
(2-dimensional) current algebras \cite{KAC}.
The (one-colour) generating 
function for fermionic partitions,
related to $\eta(x)/\eta(x^2)$,
is the same infinite product (Eq.~4) but
with $(1+x^n)^{-1}$ replacing $(1-x^n)$.
The bosonic entropy is then $\sqrt{2}$
times the fermionic entropy
and (all else being equal)
bosons take twice the energy of
fermions aymptotically \cite{POL}. 

In a heuristic spirit,
we are in effect arguing that the above
functions relate to the probability
(per unit $\log x$) of finding the
system in a particular state,
and as such 
represent structure functions
in the high energy limit,
if we assume that partons
behave like free particles at high energy.
In Figure~2a we plot the moments of $xF_3$ 
(solid circles) 
for factor-of-ten increments in $Q^2$ 
(as above)
calculated from the QCD fit,
as a function of $\ln Q^2/\Lambda^2$
on a log-log plot, ie.\ we plot $\ln M_3(N)$
versus $s = \ln \ln Q^2/\Lambda^2$
(with $\Lambda \simeq 250$ MeV) 
so that perturbative QCD evolution is reduced 
to a linear dependence with a known slope
($M_3(N) \equiv \int x^{N-1}F_3dx$ \cite{BOS}).
Extrapolating (the moments of) $xF_3$ 
to $Q^2 \sim 10^{38}$ GeV$^2$ 
($Q \sim M_{PL}$), we found to reasonable accuracy: 
$xF_3 \rightarrow [x^{\frac{1}{12}}\eta(x)/\eta(x^2)]^5$,
where the moments of this function
are represented in Figure~2a 
by the open half-circles
plotted at the largest value of $Q^2$.
This function
(clearly cloaely related to the fermionic partition function above)
is plotted versus $x$
in Figure~1a (dashed curve,
labelled $Q^2 \sim M_{{\rm PL}}^2$)
and is in fact very similar 
in general appearance 
to the function $\eta^5(x)$ (not shown),
either function
satisfying the  GLS sum-rule \cite{GLS}
for the $N=1$ moment
to an accuracy of a few percent.

The above observations
(Figure~1a/2a)
lend empirical support
to the notion that powers 
(or combinations of powers)
of $\eta(x)$ may underlie
the functional form of
the structure functions
at the highest energies 
(regarding the basic shape 
and also, it would seem in this case, 
the normalisation).
On the other hand, 
we have
no particular explanation 
for the apparent
preference for the fifth power here
(but see Ref.~\cite{SUS} below). 

The theta function $\theta_1'$ (Eq.~3)
evaluated at $z = 0$ is proportional  
to the third power of the eta function,
specifically: 
\begin{equation}
\theta_1'(0,x)/2 = \eta^3(x).
\end{equation}
where we adopt the usual convention
$\theta_i(z,x) \equiv \theta_i(z|\tau)$
(the conventional symbol $q^2$  
often used in place of  $x$,
is deliberately avoided here).
In Figure~1b we plot the evolution
of $\theta_1'$ for factor-of-ten increments in $Q^2$, 
with the a posteriori (see below)  identification
(assumed valid in the limit $z \rightarrow 0$): 
\begin{equation}
z \hspace{5mm} \simeq \hspace{5mm} {\rm const.} 
       \hspace{3mm} - \hspace{3mm} \frac{1}{2\pi} \ln \ln Q^2/\Lambda^2
          \hspace{1.0cm} (z \hspace{2mm} \simlt \hspace{2mm} 0.6) 
\end{equation}
where the constant is fixed 
such that $z=0$ corresponds to
$Q^2 \simeq 10^{38}$ GeV$^2$
(ie.\ the constant is equal
to the $Q^2$-dependent term
with $Q$ set equal to $M_{PL}$, viz.\
${\rm const.} \simeq (1/2\pi) \ln \ln M_{PL}^2/\Lambda^2$
where
$M_{PL}$ is the Planck-mass,
$M_{PL} \simeq 10^{19}$ GeV).
Comparing Figure~1a and Figure~1b,
it is apparent that the $z$-dependence
of the theta function shows a
really rather startling similarity to the
$Q^2$ dependence of the structure function,
at least over the current 
experimentally relevant range 
$Q^2 = 1 - 10^7$ GeV$^2$ 
($z \simeq 0.55-0.25$). 

A more quantitative 
comparison is achieved by plotting
(the logarithm of) the $x$-moments 
of the theta function versus $z$ (Figure~2b)
alongside the corresponding plot (Figure~2a) 
for the structure function moments
versus $s = \ln \ln Q^2/\Lambda^2$.
In each case we see a broadly linear
dependence of (the logarithm of) 
the moments on the ordinate,
with a slope which increases 
with increasing  moment number.
The theta function moments (Figure~2b) 
converge to a common value ($2\pi$)
as $z \rightarrow \pi/2$, where
$\theta_1'/2 \rightarrow 2 \pi \delta(x-1)$.
While the structure function moments cannot 
be plotted (versus $s$) for $Q^2 < \Lambda^2$
and are not plotted anyway for $Q^2 < 1$ GeV$^2$,
as explained above,
it is evident nonetheless (Figure~2a) 
that if the linear dependence 
of the QCD fit
were naively continued to 
lower $Q^2$
(broken lines) 
the structure function moments
would also apparently converge 
in very similar way
(this striking feature
of the data
has no explanation in
the context of perturbative QCD, 
and has been cited  as evidence for 
`valons' \cite{HWA},
and is also apparently a prediction 
of bag-models \cite{JAF}
and possibly also
chiral-soliton models \cite{SOL}). 
Of course the structure function
becomes proportional to $\delta(x-1)$
at very small values of $Q^2$,
as a consequence of the dominance
of elastic scattering 
($xF_3 \rightarrow 5.7 \times \delta(x-1)$ 
, as $Q^2 \rightarrow 0$ \cite{BOS}).

Assuming the correspondence Eq.~7,
in Figure~3 we compare the slopes 
(with respect to $\log Q^2$) 
of the moments of the 
theta function (solid curve)
directly with the leading-order QCD
non-singlet anomalous dimensions,
evaluated for three quark flavours
(broken histogram)
for the first ten moments ($N=1-10$).
The theta function
slopes are evaluated
at some not-too-small value of $z$
($z = \pi/6 \simeq 0.5$),
roughly where present-day 
measurements cluster.
Somewhat remarkably
the theta function slopes are seen to reproduce
the three flavour anomalous dimensions
to an accuracy of a few percent or so,
over the range of moment numbers plotted,
and it is this comparison which justifies,
(a posteriori) the identification Eq.~7, 
where the normalising factor $1/(2\pi)$
is inserted purely empirically.
At large $N$
the non-singlet anomalous dimensions grow
like $\frac{16}{27} \log N$,
while the corresponding dependence for the
theta function slopes 
appears to be  $\frac{1}{\pi}\sqrt{2N}$.
Large $N$ moments are not
explored experimentally, however.

Note that this remarkable
quantitative agreement,
whether partly (or even wholly) 
`accidental', 
extends over the entire kinematic range
explored/fitted experimentally.
At higher values of $Q^2$
($z$ $\simlt$ $\pi/12$)
the $z$-dependence of the theta function
moments (Figure~2b) begins to `flatten-out', 
with Bjorken scaling
holding `exactly' 
for the theta function
at the very highest energies
($Q \rightarrow M_{PL}$).
While this effect is naturally
not present in the QCD fit
(extrapolated to high energy)
it is clear that
such high energies 
($Q^2$ $\simgt$ $10^7$ GeV$^2$)
have so far also not been 
explored experimentally.
(Certainly in any case, QCD
can hardly be expected
to hold unmodified
for $Q \rightarrow M_{PL}$).
Taking the $Q^2$-dependence
of the theta function seriously,
at the lower values of $Q^2$
($z$ $\simgt$ $\pi/12$)
these effects are small (see above)
and better than 1\% measurements
of $xF_3$ would be required 
to establish them even at HERA,
$Q^2 \sim 10^4$ GeV$^2$ \cite{HERA}.
By the same token,
substantive deviations 
($\simgt$ 20\%)
from the familiar perturbative QCD 
$Q^2$-dependence might be expected for 
$Q^2$ $\simgt$ $10^7$ GeV$^2$
and could conceivably show-up, 
for example 
at a next-generation $ep$ collider.

To summarise,
we have seen that
the classical theta function
$\theta_1'$ bears a clear resemblance 
to a non-singlet structure function,
like $xF_3$.
While these two functions are not identical,
even in the region explored experimentally,
it is hard to believe that the 
observed similarities
are entirely accidental.
We are led to conclude
that a qualitative relationship
exists between structure functions
and theta functions,
a notion on which we will enlarge below.
This is the main conclusion of our
investigation and it rests
on the analysis of the non-singlet
structure functions, 
Figures~1-3 above.

Regarding
singlet structure functions,
ie.\ $F_2 = Q(x)+\bar{Q}(x)$,
or the closely related gluon distribution 
$G(x)$ in QCD,
we have only a few comments.
First,
we note a similarity of form
between, for example, the relation 
$xF_3 = Q(x)-\bar{Q}(x)$
and a theta function identity like
$\theta_2^4(0) = \theta_3^4(0)-\theta_4^4(0)$,
essentially the Jacobi `obscure' formula \cite{ELL}:
\begin{equation}
16x^{\frac{1}{2}}\prod_{n=1}^{\infty} (1+x^n)^8
       = \prod_{n=1}^{\infty} (1+x^{n-\frac{1}{2}})^8
            -\prod_{n=1}^{\infty} (1-x^{n-\frac{1}{2}})^8
\end{equation}
Multiplying the theta function identity
by the infinite product $\prod (1-x^n)^{24}$
and plotting the 
$x$-dependence of the various terms in Figure~4,
the resulting curves are reminiscent
of the familiar behaviour of 
$Q(x)$, $\bar{Q}(x)$ etc.\ at small $x$.
This suggests that singlet structure functions
may be related in some way to functions 
like $\theta_3$ and $\theta_4$ 
which are expressible as simple infinite products \cite{ELL},
without any powers of $x$.
In this connection,
we note further that,
exploiting the modular covarince of Eq.~4, 
the infinite product $\prod (1-x^n)$
can be shown \cite{RAM}
to be remarkably well approximated by:
\begin{equation}
\prod_{n=1}^{\infty} (1-x^n) 
   \hspace{4mm} \stackrel{x \rightarrow 1}{ \simeq} 
                \hspace{4mm} \sqrt{2\pi} 
         \frac{x^{-\frac{1}{24}}}{\sqrt{\ln \frac{1}{x}}}
             \exp (-\frac{\pi^2}{6 \ln \frac{1}{x}})
\end{equation}
(with less than 10\% deviation even
for $x$ as small as $x = 10^{-6}$
for example) 
where the right-hand side bears a clear resemblance
to the well-known
$Q^2 \rightarrow \infty$ expression 
for the (unintegrated) gluon distribution at small $x$ 
($x$ $\simlt$ $10^{-3}$) \cite{FOR}:
\begin{equation}
\frac{\partial \, G(x)}{\partial \, \ln k_T^2}
          \hspace{4mm} \stackrel{x \rightarrow 0}{\simeq} 
                  \hspace{4mm} 
              \sqrt{\frac{k_T^2}{k_1^2}} 
         \frac{x^{-\lambda_{BFKL}}}{\sqrt{\ln \frac{1}{x}}}
             \exp (-\frac{\ln^2 k_T^2/k_0^2}{56 \zeta(3) 
                \bar{\alpha} \ln \frac{1}{x}})
\end{equation}
obtained by solving the BFKL \cite{BFK} 
equation analytically \cite{CAT} 
(and it will be noted that 
despite the `formal' inconsistency
of the limits (Eq.~9 vs.\ Eq.~10)
the ranges of validity actually
overlap over three orders of magnitude in $x$). 
In Eq.~10, $\bar{\alpha} \equiv \frac{3}{\pi}\alpha$
(where $\alpha$ is the QCD gauge-coupling)
and $\lambda_{BFKL} = 4 \ln 2 \hspace{2mm} \bar{\alpha}$
($k_0^2$, $k_1^2$ are non-perturbative scale parameters).

It should be pointed out that
if the analytic similarity
of Eq.~9 to Eq.~10
is not accidental,
then there may be someting
to learn here about
how the BFKL singularity
is regulated in QCD
(Eq.~9 LHS does not obviously
violate the Froissart bound).
Likewise,
the gauge-coupling \cite{ZIC}
would seem to be related
to the index $1/24$ above
(we note that
for QCD, the Freudenthal de Vries `strange' formula
reduces to $8 \times C_A =24$, where $C_A=3$,
while for three generations of quarks
the corresponding net strength of the quark-gluon vertex
in QCD is given by $18 \times C_F = 24$, where $C_F=4/3$).
Casimir himself
postulated a relation between 
the zero-point energy coefficient
and the gauge coupling \cite{CAS2}.
In (2-dimesional) dilaton gravity,
the conformal anomaly $c/24$
is known to play the role 
of the (dimensionless) 
inverse Newton constant \cite{CAD}. 

Of course theta functions
solve the diffusion equation
$\partial^2 \theta /\partial z^2 
= (4i/ \pi) \, \partial \theta /\partial \tau$
and the link between BFKL and diffusion
has long been recognised,
having its physical origin in
the absence of transverse momentum ($k_T$)
ordering in the BFKL ladder \cite{KW}.
While the kind of global similarity
to theta functions
seen in the non-singlet case (Fig.~1-2)
would certainly suggest
that diffusion plays an important role 
for longitudial variables also,
clearly it is not possible (from perturbative QCD alone) to derive 
a double-differential equation for structure functions
(comparable to the diffusion equation) for arbitrary $x$ and $Q^2$.
In the limit $Q^2 \rightarrow \infty$, $x \rightarrow 0$
the DGLAP equation \cite{LAP} for $G(x)$ 
in fact reduces to a wave equation:
$\partial^2 G /\partial z \partial \tau
   \simeq  2 \pi \, \partial G /\partial \tau +\dots$\cite{BAL},
but it should be remembered that 
the DGLAP approach omits explicitly 
the unordered $k_T$ contribution above 
(assumes strong $k_T$ ordering).
Regarding our extrapolation
to $Q \rightarrow M_{Pl}$,
where presumably 
quantum gravity effects
cannot be neglected 
(ie.\ beyond the QCD context)
it is perhaps worth noting that
the partition function 
for a microscopic black-hole 
is known to satisfy
the 1D-diffusion equation \cite{KA}.
While the smooth extrapolation
of DIS physics to black-hole physics
implied in the present paper
should obviously be viewed with due caution,
we note further that
a partial analogy between
microscopic black-hole evaporation
(via Hawking radiation \cite{HW})
and QCD fragmentation
has in fact already been recognised 
in the literature \cite{MA}
(and see also Ref.~\cite{SUS} below).

If anything, 
the above observations
on singlet structure functions
reinforce the case for
a connection between 
structure functions
and theta functions,
which was the main conclusion
of the non-singlet analysis.
That is not to say that
we have in hand a complete and quantitative
description of
the neutrino-nucleon structure functions
in terms of the Jacobi theta functions.
The comparisons we have made
are at best
only partially successful
and sometimes
apparently self-contradictory
(eg.\ $xF_3$ cannot `equal' $\theta_1'$
which behaves like $x^{\frac{1}{8}}$
at small $x$ and simultaneously display 
an $x^{\frac{1}{2}}$ dependence
as implied by Eq~8).
Also some generalisation of Eq.~7,
to cover the low $Q^2$ region,
needs to be found, 
which may be difficult in practice.
Nonetheless, 
considering the non-singlet
and singlet cases together,
the accumulated evidence for 
a qualitative (or `formal') relation 
between structure functions
and theta functions 
is undeniably rather strong.
The most obvious possibility
would be that structure functions
{\em are} theta functions of some sort
(or powers of theta functions)
or some variant of 
theta functions with 
similar behaviour. 
Possibly, theta functions 
relate to some
`idealised' structure function 
eg.\ of a pure-gauge system
(or in the massless/SUSY limit)
unobservable experimentally.
The theoretical derivation of such a result
might conceivably proceed within
the representation theory
of some infinite dimensional algebra \cite{KAC}.

To conclude,
our point of view
is that theta functions
arise in connection with
the highest energy scales,
eg.\ the Planck-scale
or the unification scale
(if our observations mean anything at all,
{\em some} high energy scale {\em must} be associated
with the $z=0$ for the theta functions, 
and the Planck/unification scale 
seems the obvious best guess, 
cf.\ Eq.~7).
While
in the QCD context 
we are often used 
to evolving models/fits 
from lower to higher energies,
one might reasonably argue that
symmetry and simplicity
are in fact more naturally associated
with the highest energy scales,
so that ultimately,
evolving 
functional forms {\em down} 
in energy may come 
to seem much more sensible.
We remark once more
that modular functions
and modular symmetry
already
feature prominently 
in theoretical considerations of
Planck-scale physics.
For example, string theory 
threshold corrections
to the reciprocal gauge-couplings \cite{DIX}
turn out to be in part
proportional to $\ln | \eta(\tau)  |^4$,
(with $\tau$ the value of a 
given moduli field
scaled to the Planck-energy).
Also,
in five-dimensional D-brane theory
the partition function
of a microscopic black-hole
(small with respect to the compactification radii)
can be expressed in terms
of integral powers
of the eta-function \cite{SUS}.
The possibility that such apparently
abstract theoretical concerns
may already be mirrored
in existing phenomenology
(cf.\ $xF_3 \rightarrow \; \sim \! \! \eta^5(x)$ 
for $Q \rightarrow M_{PL}$)
is both exciting and 
potentially important.

For the future,
emerging naturally from this analysis
large-$x$/small-$x$ `duality' relations
for structure functions
of the form Eq.~1 (or similar)
are certainly an attractive 
and ultimately testable possibility.
While the non-singlet case
would seem to offer
the best chance,
empirical fits to singlet
structure functions
(eg.\ to $F_2^{ep}$  Ref.~\cite{HAIDT})
do in fact 
suggest `critical' x-values
of a sort,
apparently $Q^2$-independent
(with $x_0 \sim \exp(-\pi)$).
With {\em some} correlation between 
the large-$x$ and small-$x$ behaviour
of structure functions undoubtedly
physically plausible \cite{FEYN},
we have every reason to hope that
definitive
experimental or theoretical evidence
for the existence of
relations of the above form
will eventually be found.
In this last regard, 
we are pleased to call attention to the
work of R. Janik \cite{JANIK},
pointing to a connection
between $xF_3$ and 
the classical elliptic theta functions,
established 
(for transverse variables)
in the Regge limit of QCD,
via the modular invariance 
of the `odderon'.

\vspace{5mm}
\noindent {\bf Acknowledgement}

\noindent It is a pleasure to thank R. G. Roberts
          for helpful discussions and encouragement.

\newpage

\noindent {\bf {\large Figure Captions}}

\vspace{10mm}
\noindent Figure~1 
a) The $x$-dependence 
of the non-singlet neutrino-nucleon 
structure function $xF_3$,
based on the standard MRS \cite{MRS} 
QCD fit to existing data,
for factor-of-ten increments in $Q^2$,
$Q^2 = 1 - 10^7$ GeV$^2$ (solid curves).
The dashed curve labelled
$Q^2 \sim M_{{\rm PL}}^2$ is the function
$[x^{\frac{1}{12}}\eta(x)/\eta(x^2)]^5$
(where $\eta(x)$ is the Dedekind 
eta function \cite{KAC} Eq.~4)
which provides a reasonably accurate
representation of the QCD fit extrapolated
to $Q^2 \sim 10^{38}$ GeV$^2$ 
(see Figure 2a).
b) The theta function $\theta_1'(z,x)/2$
plotted as a function of $x$ (solid curves) 
for a sequence of $z$-values,
$z = 0.55-0.25$ (see Eq.~7).
The limiting form of $\theta_1'/2$ 
at $z = 0$ is $\eta^3(x)$ (dashed curve).

\vspace{10mm}
\noindent Figure~2 
a) (The natural logarithm of) 
the $x$-moments ($N=1-7$) of 
the structure function $xF_3$,
plotted as a function of 
$s = \ln \ln Q^2/\Lambda^2$,
for $Q^2 = 1-10^7$ GeV$^2$
(filled points).
Extrapolating the QCD 
dependence (solid lines)
to $Q^2 \sim 10^{38}$ GeV$^2$,
the extrapolated moments are seen
to be reasonably accurately reproduced
by the moments of the function
$[x^{\frac{1}{12}}\eta(x)/\eta(x^2)]^5$
(open half-points).
A naive extrapolation
to low $Q^2$ (broken lines) 
shows the moments apparently 
converging (see text).
b) The $x$-moments of the theta function 
($\theta_1'/2$) plotted as a function of $z$.
The theta function moments also converge
to a common value as $z \rightarrow \pi/2$
(ie.\ at small values of $Q^2$)
where $\theta_1'/2 \rightarrow 2 \pi \delta(x-1)$.
At small $z$ (ie.\ at large values of $Q^2$) 
the $z$-dependence `flattens-off',
so that Bjorken scaling holds
exactly  for $\theta_1'$ 
as $z \rightarrow 0$.

\vspace{10mm}
\noindent Figure~3
The slopes of the theta function moments
with respect to $\log Q^2$, 
(assuming the correspondence Eq.~7)
evaluated at $z=\pi/6 \simeq 0.5$,
plotted as a function of 
moment number $N$ (solid curve).
The broken histogram shows
the leading-order QCD anomalous dimensions,
calculated for three quark flavours.
The theta function slopes reproduce the
leading-order QCD anomalous dimensions
(to an accuracy of a few percent or so)
over the range of moment numbers plotted.

\vspace{10mm}
\noindent Figure~4 
Terms in the theta function identity
$\theta_2^4(0)=\theta_3^4(0)-\theta_4^4(0)$ \cite{ELL} 
(essentially the Jacobi `obscure' formula Eq.~8)
multiplied by the infinite product
$\prod (1-x^n)^{24}$, plotted
as a function of $x$,
for small-$x$.
The resulting curves are reminiscent
of the familiar behaviour of 
$Q(x)$, $\bar{Q}(x)$ etc.\
at small-$x$
(cf.\
$\theta_2^4(0)=\theta_3^4(0)-\theta_4^4(0)$
: $xF_3 = Q(x)-\bar{Q}(x)$),
suggesting that singlet structure functions
may be related to functions like 
$\theta_3$ and $\theta_4$,
expressible as simple infinite products,
without any powers of $x$.

\newpage
\pagestyle{empty}
\begin{figure*}[hbt]
\epsfig{figure=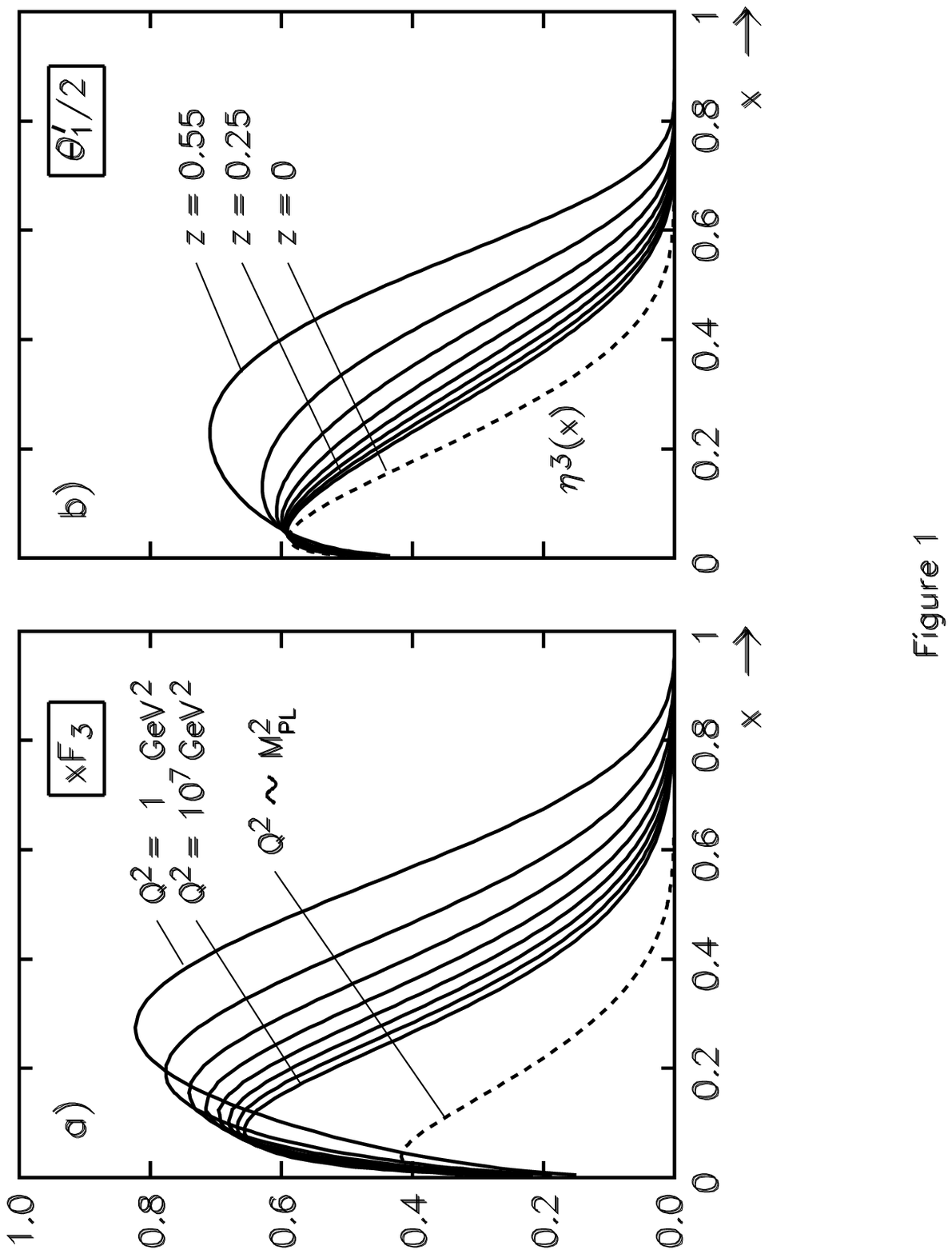,width=160mm,bbllx=100pt,bblly=120pt
,bburx=550pt,bbury=720pt}
\end{figure*}
\newpage
\begin{figure*}[hbt]
\epsfig{figure=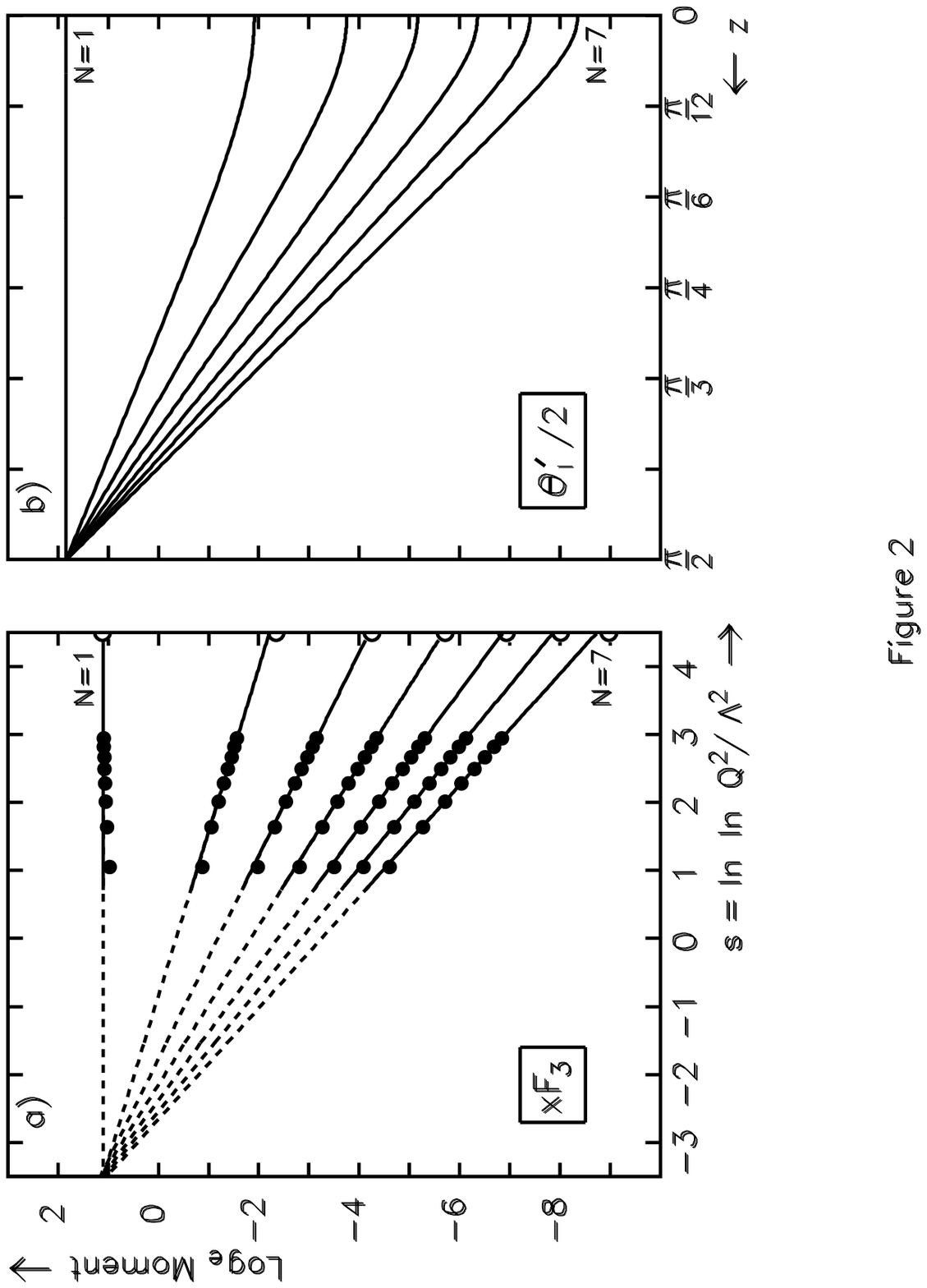,width=160mm,bbllx=100pt,bblly=120pt
,bburx=550pt,bbury=720pt}
\end{figure*}
\newpage
\begin{figure*}[hbt]
\epsfig{figure=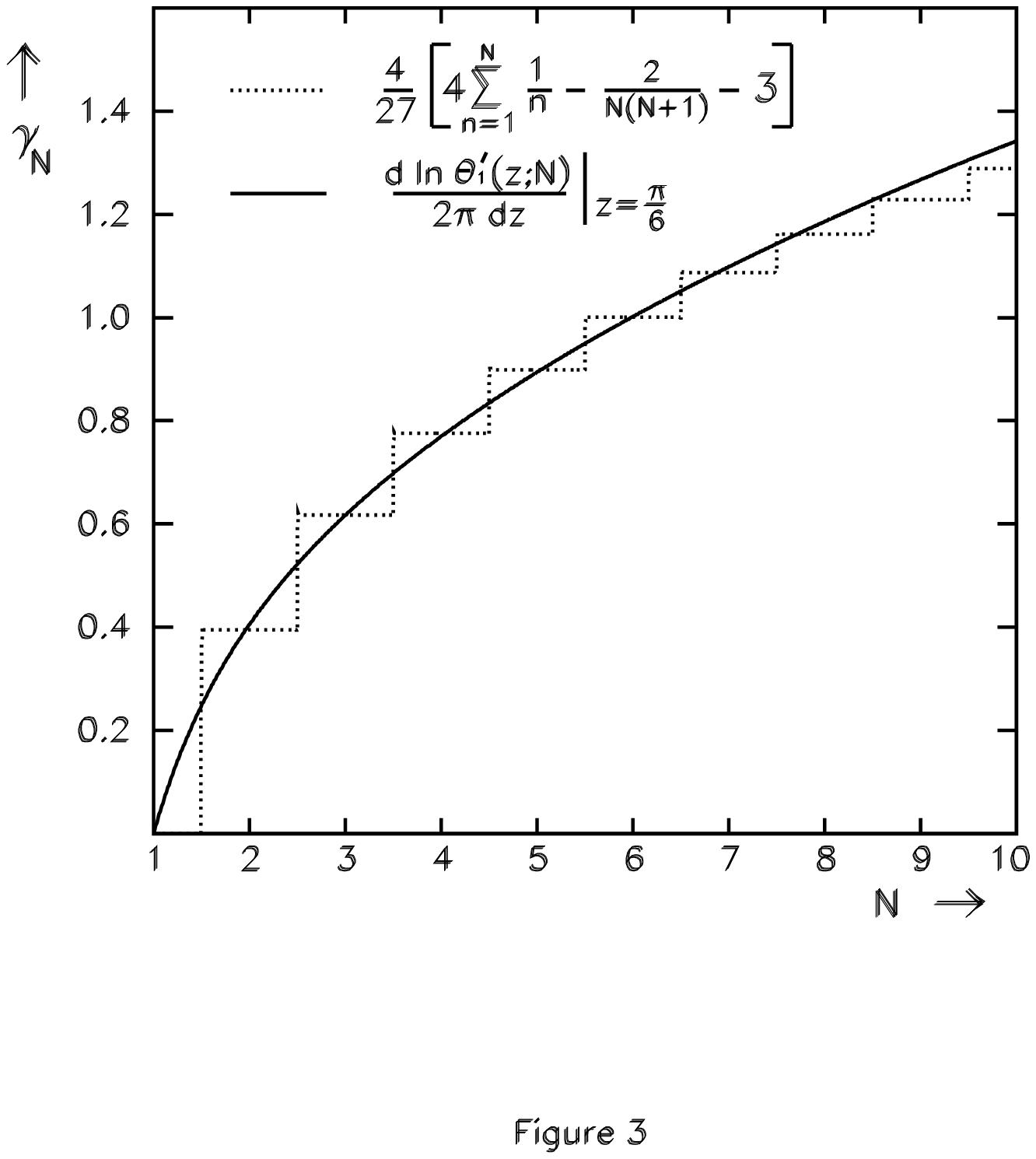,width=150mm,bbllx=80pt,bblly=100pt
,bburx=530pt,bbury=700pt}
\end{figure*}
\newpage
\begin{figure*}[hbt]
\epsfig{figure=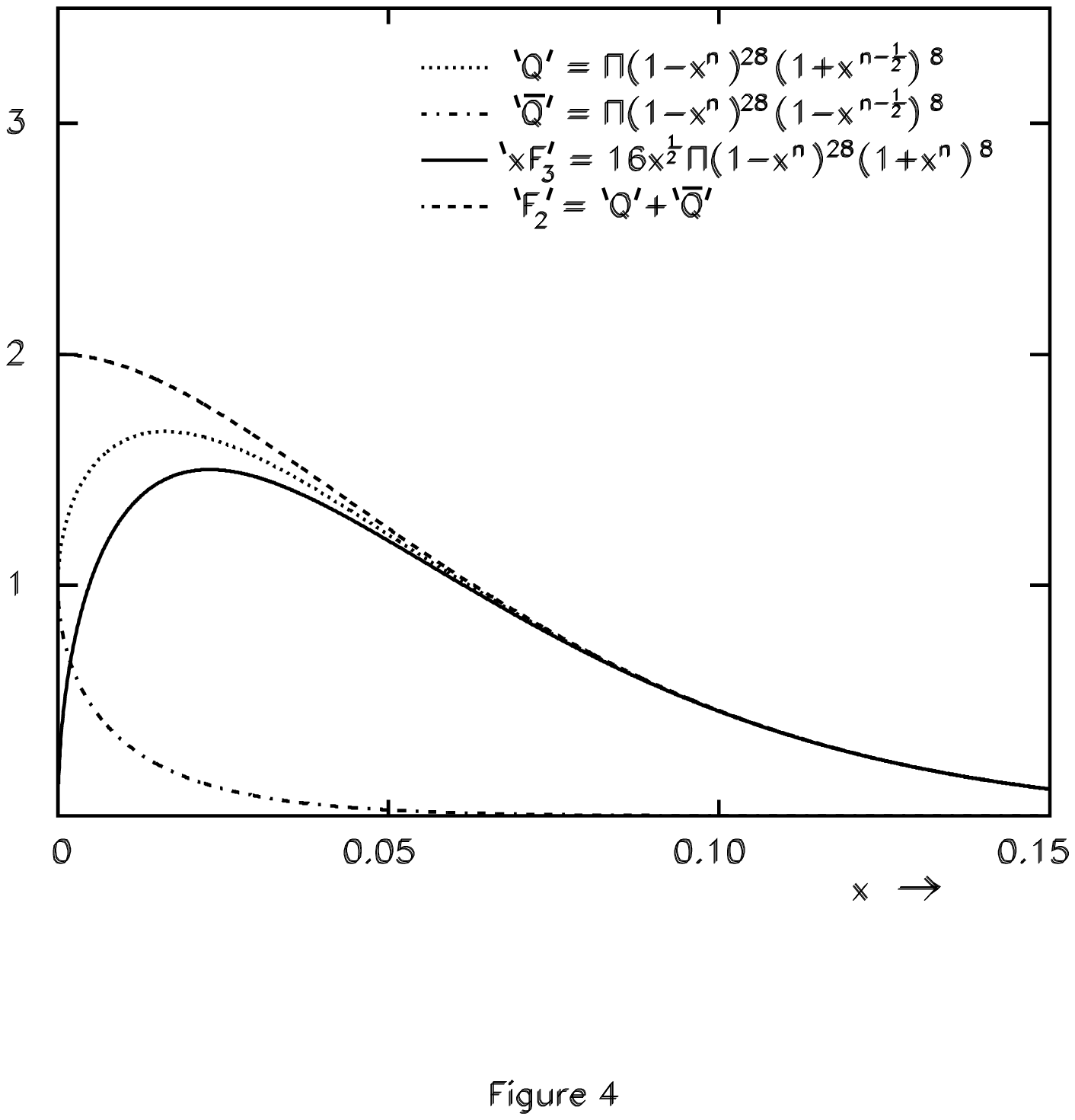,width=150mm,bbllx=85pt,bblly=100pt
,bburx=535pt,bbury=700pt}
\end{figure*}
       
\end{document}